\documentclass[nofootinbib,aps,pra,showpacs,superscriptaddress,preprint]{revtex4}
\usepackage{amsmath}
\usepackage{subfig}
\usepackage{graphicx}

\catcode`ð=\active
\defð{\u{g}}
\catcode`Ð=\active
\defÐ{\u{G}}
\catcode`Ý=\active
\defÝ{\. I}
\catcode`ö=\active
\defö{\"{o}}
\catcode`Ö=\active
\defÖ{\"O}
\catcode`ü=\active
\defü{\"{u}}
\catcode`Ü=\active
\defÜ{\"{U}}
\catcode`Þ=\active
\defÞ{\c{S}}
\catcode`þ=\active
\defþ{\c{s}}
\catcode`ý=\active
\defý{{\i}}
\catcode`ç=\active
\defç{\d{c}}
\catcode`Ç=\active
\defÇ{\d{C}}

\begin{document}

\title{Approximate Analytical Solutions of a Two-Term Diatomic
Molecular Potential with Centrifugal Barrier}
\author{\small Altuð Arda}
\email[E-mail: ]{arda@hacettepe.edu.tr}\affiliation{Department of
Physics Education, Hacettepe University, 06800, Ankara,Turkey}
\author{\small Ramazan Sever}
\email[E-mail: ]{sever@metu.edu.tr}\affiliation{Department of
Physics, Middle East Technical  University, 06531, Ankara,Turkey}

\begin{abstract}
Approximate analytical bound state solutions of the radial
Schrödinger equation are studied for a two-term diatomic molecular
potential in terms of the hypergeometric functions for the cases
where $q\geq1$ and $q=0$. The energy eigenvalues and the
corresponding normalized wave functions of the Manning-Rosen
potential, the 'standard' Hulth\'{e}n potential and the
generalized Morse potential are briefly studied as special cases. It is observed that our analytical results are the same with the ones obtained before.
\end{abstract}

\maketitle

\newpage

\section{Introduction}
In this letter, we study the bound state solutions of a diatomic
molecular potential having the form
\begin{eqnarray}
V(r,\beta,q)=-V_{0}\,\frac{e^{-\beta r}}{1-qe^{-\beta
r}}+V_{1}\,\frac{e^{-2\beta r}}{(1-qe^{-\beta r})^2}\,,
\end{eqnarray}
which has been firstly presented by Sun to fit experimental data
of some diatomic molecular systems [1]. Analytical study of the
above potential could be interesting since it involves several
potential forms (for example, $q=0$ gives the Morse potential,
$q>0$ corresponds to the 'generalized' Hulth\'{e}n potential,
\textit{etc.}) meaning that we can simply extend the solutions to
the ones of these special cases.

The above potential is one of the central potentials which are a
powerful ground for experimental and theoretical computations in
different areas of physics such as in high energy physics where
they were used to describe hadrons as bound states [2], in atomic
physics where some important subjects such as binding energy and
inclusive momentum distributions are studied by using of central
potentials [3], in theoretical molecular dynamics model to study
the intramolecular and intermolecular interactions and atomic pair
correlation functions [4]. Moreover, the central potentials have
been used in an important quantum mechanical problem which is also
related with quantum information theory, the Fisher uncertainty
relation and applied to the hydrogen atom and isotropic harmonic
oscillator [5] and also for some theoretical calculations within
the information theory to study some statistical quantities such
as the Boltzmann-Shannon entropy [6]. The construction of an
algorithm could be an interesting problem where the aim is to
solve the radial Schrödinger equation (SE) for a given central
potential $V(r)$ numerically [7].

To our knowledge, the potential under consideration has been
studied within the supersymmetric quantum mechanics [8] and in
terms of Green's function [9] in non-relativistic domain. We
search the bound state spectrum and the wave functions of the
above potential by using an approximation instead of the
centrifugal term in the same domain. We find an analytical
expression for the energy spectrum and obtain the normalization
constant by using some properties of the hypergeometric functions.
Throughout this work, we restrict ourself to the cases where
$q\geq1$ and $q=0$ and give our numerical results for two diatomic
molecules for different values of quantum number pair ($n, \ell$).
\section{Energy Spectrum and Wave Functions}
The radial Schrödinger equation is written [10]
\begin{eqnarray}
\frac{d^{2}R(r)}{dr^2}+\left[\frac{2m}{\hbar^2}\left[E_{n\ell}-V(r)\right]-\frac{\ell(\ell+1)}{r^2}\right]R(r)=0\,,
\end{eqnarray}
where $\ell$ is the angular momentum quantum number, $m$ is the
particle mass, $V(r)$ is the central potential and $E_{n\ell}$ is
the non-relativistic energy. Inserting Eq. (1) into Eq. (2) gives
\begin{eqnarray}
\frac{d^{2}R(r)}{dr^2}+\left[-\frac{2m
V_{1}}{\hbar^2}\,\frac{1}{(e^{\beta r}-q)^2}+\frac{2m
V_{0}}{\hbar^2}\,\frac{1}{e^{\beta r}-q}+\frac{2m
E_{n\ell}}{\hbar^2}-\frac{\ell(\ell+1)}{r^2}\right]R(r)=0\,.
\end{eqnarray}
where $V_{0}, V_{1}, \beta$ and $q$ are real parameters defined by
$V_{1}=D_{0}(e^{\mu}-q)$, $V_{0}=2V_{1}$, $\beta=\mu/r_{0}$, where
$D_{0}$ is the depth of the potential, $r_{0}$ is the equilibrium
of the molecule and $q$ is the shape parameter.

We use the following approximation [11] instead of the centrifugal
term among the others [12-15] to obtain an analytical solution of
Eq. (3)
\begin{eqnarray}
\frac{1}{r^2} \approx \beta^2\,\frac{e^{\beta r}}{(e^{\beta
r}-q)^2}\,,
\end{eqnarray}
Defining a new variable $z=qe^{-\beta r}$ and taking a trial
function as $R(z)=z^{A_{1}}(1-z)^{A_{2}}\phi(z)$ and with the help
of Eq. (4), Eq. (3) turns into
\begin{eqnarray}
z(1-z)\frac{d^2\phi(z)}{dz^2}&+&\left[1+2A_{1}-(1+2A_{1}+2A_{2})z\right]\frac{d\phi(z)}{dz}\nonumber\\
&+&\left[-2A_{1}A_{2}-A^2_{2}+\frac{2mV_{1}}{\beta^2\hbar^2}+\frac{2mV_{0}}{q\beta^2\hbar^2}\right]\phi(z)=0\,,
\end{eqnarray}
where we set the parameters
\begin{subequations}
\begin{align}
&A^2_{1}=-\frac{2m E_{n\ell}}{\hbar^2}\,,\\
&A_{2}(A_{2}-1)=\frac{1}{q}\,\ell(\ell+1)+\frac{2mV_{1}}{\beta^2\hbar^2}\,.
\end{align}
\end{subequations}

By using the abbreviations
\begin{subequations}
\begin{align}
&c=1+2A_{1}\,,\\
&b=A_{1}+A_{2}+\sqrt{A^2_{1}+\frac{2m}{\beta^2\hbar^2}\left(V_{1}+\frac{V_{0}}{q}\right)\,}\,,\\
&a=A_{1}+A_{2}-\sqrt{A^2_{1}+\frac{2m}{\beta^2\hbar^2}\left(V_{1}+\frac{V_{0}}{q}\right)\,}\,,
\end{align}
\end{subequations}
Eq. (5) becomes an equation having the form of the
hypergeometric-type equation [16]
\begin{eqnarray}
z(1-z)\phi''(z)+[c-(a+b+1)z]\phi'(z)-ab\phi(z)=0\,,
\end{eqnarray}
whose solution is the hypergeometric functions
\begin{eqnarray}
\phi(z) \sim\, _{2}F_{1}(a, b; c; z)\,.
\end{eqnarray}
When either $a$ or $b$ equals to a negative integer $-n$, the
hypergeometric function $\phi(z)$ can be reduced to a finite
solution. This gives us a polynomial of degree $n$ in Eq. (9) and
the following quantum condition
\begin{eqnarray}
A_{1}+A_{2}-\sqrt{A^2_{1}+\frac{2m}{\beta^2\hbar^2}\left(V_{1}+\frac{V_{0}}{q}\right)\,}=-n\,,
\end{eqnarray}
which gives the energy values of the two-term potential for any
$\ell$-values
\begin{eqnarray}
E_{n\ell}=-\frac{\beta^2\hbar^2}{2m}\left[\frac{n^2+(2n+1)\left(A'_{2}+\frac{1}{2}\right)+\frac{1}{q}
\left[\ell(\ell+1)-\frac{2mV_{0}}{\beta^2\hbar^2}\right]}{2n+1+2A'_{2}}\right]^2\,,
\end{eqnarray}
where
\begin{eqnarray}
A'_{2}=\sqrt{\frac{1}{4}+\frac{\ell(\ell+1)}{q}+\frac{2mV_{1}}{\beta^2\hbar^2}\,}\,.
\end{eqnarray}
By using Eq. (10) we obtain the total wave functions
\begin{eqnarray}
R(z)=Nz^{A_{1}}(1-z)^{A_{2}}\,_{2}F_{1}(-n,n+2A_{1}+2A_{2};1+2A_{1};z)\,.
\end{eqnarray}
where $N$ is the normalization constant and will be derived in
Appendix A.

We summarize our numerical results in Table 1 and 2 where the
computations are made for two diatomic molecules, namely $H_{2}$
and $LiH$. The values of potential parameters we used for these
molecules are as follows [13]: $D_{0}=4.744600$ eV,
$r_{0}=0.741600\,\AA$, $m=0.503910$ amu, $\mu=1.440558$ and
$E_{0}=\hbar^2/(mr^2_{0})=1.508343932 \times 10^{-2}$ eV for
$H_{2}$ molecule and $D_{0}=2.515287$ eV, $r_{0}=1.595600\,\AA$,
$m=0.8801221$ amu, $\mu=1.7998368$ and $E_{0}=1.865528199 \times
10^{-3}$ eV for $LiH$ molecule [17]. It is seen that the energy
values decrease while the values of the quantum numbers increase
and the energy eigenvalues are also inversely proportional with
the shape parameter for each of the molecules.

Now we intend briefly to study some special cases whose energy
eigenvalue equation obtained from Eq. (11) by suitable choices of
the potential parameters.
\subsubsection{Manning-Rosen Potential}
The Manning-Rosen potential can be written as [18]
\begin{eqnarray}
V(r)=-\frac{A\hbar^2}{2mb^2}\,\frac{1}{e^{r/b}-1}+\frac{\alpha(\alpha-1)\hbar^2}{2mb^2}\,\frac{1}{(e^{r/b}-1)^2}\,,
\end{eqnarray}
If we write our parameters as $V_{0}=\frac{A}{2b^2}$;
$V_{1}=\frac{\alpha(\alpha-1)}{2b^2}$;$\beta=\frac{1}{b}$ and
$q=1$ then we obtain the energy eigenvalues of the Manning-Rosen
potential
\begin{eqnarray}
E_{n\ell}=-\frac{1}{2b^2}\left[\frac{n^2+(2n+1)\left(\frac{1}{2}+\sqrt{\frac{1}{4}+\ell(\ell+1)+\alpha(\alpha-1)\,}\,\right)+\ell(\ell+1)-A}
{2n+1+2\sqrt{\frac{1}{4}+\ell(\ell+1)+\alpha(\alpha-1)\,}}\right]^2\,.
\end{eqnarray}
which is the same result obtained in Ref. [19]. The normalization
constant in Eq. (13) is obtained from Eq. (A8) as
\begin{eqnarray}
N=\left[\frac{1}{g(A_{1}^{(1)},A_{2}^{(1)},k)g(A_{1}^{(1)},A_{2}^{(1)},l)\,_{2}F_{1}(-2A_{2}^{(1)},1+2A_{1}^{(1)}+k+l;2+2A_{2}^{(1)}+k+l;1)}\right]^{1/2}\,.\nonumber\\
\end{eqnarray}
where $A_{1}^{(1)}=\sqrt{-2mE_{n\ell}b^2/\hbar^2\,}$ and
$A_{2}^{(1)}=(1/2)\left(1+\sqrt{1+4\ell(\ell+1)+4m\alpha(\alpha-1)/\hbar^2\,}\right)$.

We summarize our numerical results obtained from Eq. (15) in Table
3 where we set the parameters as $A=2b$ and $\alpha=0.75$ to
compare the results with the ones given in Ref. [19]. Please note
that the parameter $D_{0}$ used in Ref. [19] is zero in the
present work since our approximation used for the centrifugal term
is different from the one used in Ref. [19] where energy
eigenvalues are computed in atomic units.
\subsubsection{Standard Hulth\'{e}n Potential}
Eq. (1) gives the standard Hulth\'{e}n potential for $V_{1}=0$ and
$q=1$
\begin{eqnarray}
V(r)=-V_{0}\,\frac{e^{-\beta r}}{1-e^{-\beta r}}\,,
\end{eqnarray}
and we obtain the energy eigenvalues from Eq. (11)
\begin{eqnarray}
E_{n\ell}=-\frac{\beta^2\hbar^2}{2m}\left[\frac{(n+\ell)(n+\ell+2)+1-\frac{2mV_{0}}{\beta^2\hbar^2}}{2(n+1+\ell)}\right]^2\,.
\end{eqnarray}
and the normalization constant of the corresponding wave functions
from Eq. (A8)
\begin{eqnarray}
N=\left[\frac{\Gamma(2+4A_{2}^{(2)}+k+\ell)}{g(A_{1},A_{2}^{(2)},k)g(A_{1},A_{2}^{(2)},\ell)(1+2A_{2}^{(2)}+k+\ell)\Gamma(1-2A_{1}+4A_{2}^{(2)})}\right]^{1/2}
\end{eqnarray}
where $A_{2}^{(2)}=1+\ell$\,. Choosing the parameters as
$\beta=\frac{1}{a}$ and $V_{0}=\alpha$ gives the following
expression ($m=\hbar=1$)
\begin{eqnarray}
E_{n\ell}=-\frac{1}{2a^2}\left[\frac{(n+\ell)(n+\ell+2)+1-2\alpha
a^2}{2(n+1+\ell)}\right]^2\,.
\end{eqnarray}
which is the same result given in Ref. [20]. The standard
Hulth\'{e}n potential in Eq. (16) could gives the Coulomb
potential for $\beta r \ll 1$
\begin{eqnarray}
V(r)=-\frac{Ze^2}{r}\,,
\end{eqnarray}
where we set $V_{0}=Ze^2\beta$\,. We obtain the energy spectrum of
the Coulomb potential from Eq. (18) ($m=\hbar=e=1$)
\begin{eqnarray}
E_{n\ell}=-\frac{Z^2}{2(n+1+\ell)^2}\,.
\end{eqnarray}
which is the same result obtained in Ref. [19]. The normalization
constant of the corresponding wave functions is given with the
help of Eq. (19) under the above assumptions.

The numerical energy values of the Hulth\'{e}n potential obtained
from Eq. (18) are placed in second part of Table 3 for different
quantum number pair ($n, \ell$) in atomic units. We choose the
parameters as $V_{0}=\beta=\delta$ as in Ref [21].
\subsubsection{Generalized Morse Potential}
We obtain the generalized Morse potential for the limit $q
\rightarrow 0$ in Eq. (1)
\begin{eqnarray}
V(r)=V_{1}e^{-2\beta r}-V_{0}e^{-\beta r}\,,
\end{eqnarray}
which gives the following equation for $s$-waves
\begin{eqnarray}
\left(\frac{d^2}{dr^2}-\frac{2mV_{1}}{\hbar^2}\,e^{-\beta
r}+\frac{2mV_{0}}{\hbar^2}\,e^{-2\beta
r}+\frac{2mE_{n\ell}}{\hbar^2}\right)R(r)=0\,,
\end{eqnarray}
Defining a new variable $z=e^{-\beta r}$ and taking the wave
function of the form $R(z)=e^{-B_{1}z/2}z^{B_{2}/2}\phi(z)$, we
obtain
\begin{eqnarray}
z\frac{d^2\phi(z)}{dz^2}+\left(1+B_{2}-B_{1}z\right)\frac{d\phi(z)}{dz}
+\left[-\frac{B_{1}B_{2}}{2}-\frac{B_{1}}{2}+\frac{2mV_{0}}{\beta^2\hbar^2}\right]\phi(z)=0\,,
\end{eqnarray}
where $B^2_{1}=\frac{8mV_{0}}{\beta^2\hbar^2}$ and
$B^2_{2}=-\frac{8mE}{\beta^2\hbar^2}$. Using a new variable
$y=B_{1}z$ gives
\begin{eqnarray}
y\frac{d^2\phi(y)}{dy^2}+\left(1+B_{2}-y\right)\frac{d\phi(z)}{dz}
+\left[-\frac{B_{2}}{2}-\frac{1}{2}+\frac{2mV_{0}}{B_{1}\beta^2\hbar^2}\right]\phi(y)=0\,,
\end{eqnarray}
which is the Laguerre differential equation
\begin{eqnarray}
xy''+(\alpha+1-y)y'+ny=0\,.
\end{eqnarray}
where the factor $n$ should be zero or a positive integer to get a
polynomial solution [22]. So, the solution of Eq. (25) are given
in terms of the Laguerre polynomials as
\begin{eqnarray}
\phi(y) \sim L_{n}^{\overline{\sigma}}(y)\,,
\end{eqnarray}
where $\overline{\sigma}=B_{2}$ and
$n=-\frac{B_{2}}{2}-\frac{1}{2}+\frac{2mV_{0}}{B_{1}\beta^2\hbar^2}$\,.
We get the total eigenfunctions of the Morse potential
\begin{eqnarray}
R(z)=Ne^{-B_{1}z/2}z^{B_{2}/2}L_{n}^{\overline{\sigma}}(B_{1}z)\,.
\end{eqnarray}
and the energy eigenvalues
\begin{eqnarray}
E_{n\ell}=-\frac{\beta^2\hbar^2}{8m}\left\{2n+1-\frac{V_{0}}{\beta\hbar}\sqrt{\frac{2m}{V_{1}}\,}\right\}^2\,.
\end{eqnarray}
We present the numerical energy values of the Morse potential
obtained from the above equation in Table 3. We give the results
for $H_{2}$ molecule (in $eV$) by taking the same parameter values
to obtain the results given in Table 1 and by setting the
potential parameters as $V_{0}=2D_{0}$, $V_{1}=D_{0}$.

Using the following representation of the Laguerre polynomials
[22]
\begin{eqnarray}
L_{n}^{\overline{\sigma}}(x)=\sum_{k=0}^{n}(-1)^{k}\Bigg(\begin{array}{c}
  n+\overline{\sigma} \\
  n-k
\end{array}\Bigg)\frac{x^{k}}{k!}\,,
\end{eqnarray}
the normalization condition is written as
\begin{eqnarray}
\left|N\right|^{2}g(n)g(m)\int_{0}^{1}e^{B_{1}z}z^{B_{2}+2k}dz=1\,,
\end{eqnarray}
where
\begin{eqnarray}
g(n)=\sum_{k=0}^{n}(-1)^{k}\Bigg(\begin{array}{c}
  n+B_{2} \\
  n-k
\end{array}\Bigg)\frac{B_{1}^{k}}{k!}\,\,;g(m)=\sum_{k=0}^{n}(-1)^{k}\Bigg(\begin{array}{c}
  m+B_{2} \\
  m-k
\end{array}\Bigg)\frac{B_{1}^{k}}{k!}\,.
\end{eqnarray}
Changing the variable $t=B_{1}z$ in Eq. (32) gives
\begin{eqnarray}
\left|N\right|^{2}g(n)g(m)B_{1}^{-(1+B_{2}+2k)}\int_{0}^{1}e^{-t}t^{B_{2}+2k}dt=1\,,
\end{eqnarray}
which includes the incomplete Gamma function defined as [22]
\begin{eqnarray}
\gamma(a,x)\equiv\int_{0}^{x}t^{a-1}e^{-t}dt=\frac{1}{a}x^{a}e^{-x}\,_{1}F_{1}(1;1+a;x)\,,
\end{eqnarray}
Finally the normalization constant is obtained as
\begin{eqnarray}
N=\left[\frac{e\Omega
B_{1}^{\Omega}}{g(n)g(m)\,_{1}F_{1}(1;1+\Omega;1)}\right]^{1/2}\,,\,\,\,\Omega=1+B_{2}+2k\,.
\end{eqnarray}

\section{Conclusions}

We have studied the approximate bound state solutions of the
radial SE equation for a two-term potential. We have obtained the
energy eigenvalues and the corresponding normalized wave functions
approximately in terms of the hypergeometric functions. We have
presented our numerical results of the energy eigenvalues of two
diatomic molecules in Tables 1 and 2. We have also studied the
analytical bound state solutions of the Manning-Rosen potential,
the 'standard' Hulth\'{e}n potential and the generalized Morse
potential as special cases. We have observed that our all
analytical results are the same with the ones obtained in the
literature. We have also summarized some numerical results of the
energy eigenvalues of the above three potentials in Table 3 and
observed that our results are good agreement with the ones
obtained before.

\section{Acknowledgments}
This research was partially supported by the Scientific and
Technical Research Council of Turkey.

\appendix

\section{Normalization Constant}
The wave functions in Eq. (13) is
\begin{eqnarray}
R(z)=Nz^{A_{1}}(1-z)^{A_{2}}\,_{2}F_{1}(-n,n+2A_{1}+2A_{2};1+2A_{1};z)\,,
\end{eqnarray}
which is written in terms of the new variable $z=q\xi$
($0\leq\xi\leq1$)
\begin{eqnarray}
R(q\xi)=Nq^{A_{1}}\xi^{A_{1}}(1-q\xi)^{A_{2}}\,_{2}F_{1}(-n,n+2A_{1}+2A_{2};1+2A_{1};q\xi)\,,
\end{eqnarray}
The normalization condition
$\int_{0}^{1}\left|R(q\xi)\right|^2d\xi=1$ gives
\begin{eqnarray}
\left|N\right|^2q^{1+2A_{1}}\int_{0}^{1}\xi^{2A_{1}}(1-q\xi)^{2A_{2}}\left[\,_{2}F_{1}(-n,n+2A_{1}+2A_{2};1+2A_{1};q\xi)\right]^2dx=1\,.
\end{eqnarray}
Using the representation of the hypergeometric functions [22]
\begin{eqnarray}
\,_{2}F_{1}(-n,b;c;z)=\sum_{k=0}^{n}\frac{(-n)_{k}(b)_{k}}{(c)_{k}k!}z^{k}\,,
\end{eqnarray}
Eq. (A3) becomes
\begin{eqnarray}
\left|N\right|^2q^{1+2A_{1}}g(A_{1},A_{2},k)g(A_{1},A_{2},l)\int_{0}^{1}\xi^{2A_{1}+k+l}(1-q\xi)^{2A_{2}}d\xi=1\,,
\end{eqnarray}
where
$(-n)_{k}=(-1)^{k}(n-k+1)_{k}=(-1)^{k}\,\frac{\Gamma(n+1)}{\Gamma(n-k+1)}$
and
\begin{eqnarray}
g(A_{1},A_{2},k)=\sum_{k=0}^{n}\frac{(-n)_{k}(n+2A_{1}+2A_{2})_{k}}{(1+2A_{1})_{k}k!}z^{k}\,.
\end{eqnarray}
and $g(A_{1},A_{2},l)=g(A_{1},A_{2},k \rightarrow l)$\,.

Using the following identity  for the hypergeometric functions
[22]
\begin{eqnarray}
\,_{2}F_{1}(\alpha',\beta';\delta';z)=\frac{\Gamma(\delta')}{\Gamma(\beta')\Gamma(\delta'-\beta')}\,\int_{0}^{1}t^{\beta'-1}
(1-t)^{\delta'-\beta'-1}(1-tz)^{-\alpha'}dt\,,
\end{eqnarray}
we obtain the normalization constant from Eq. (A5)
\begin{eqnarray}
N=\left[\frac{1}{q^{1+2A_{1}}g(A_{1},A_{2},k)g(A_{1},A_{2},l)\,_{2}F_{1}(-2A_{2},1+2A_{1}+k+l;2+2A_{2}+k+l;q)}\right]^{1/2}\,.\nonumber\\
\end{eqnarray}

\newpage

\begin{table}
\begin{ruledtabular}
\caption{Energy eigenvalues of the $H_{2}$ molecule in $eV$
($E_{n\ell}<0$).}
\begin{tabular}{cccc}
$n\,\,\,\,\,\,\ell$ & $q=1.25$ & $q=1.50$ & $q=1.75$ \\
\hline
0\,\,\,\,\,0 & 3.8099200 & 2.3409000 & 1.5021100 \\
1\,\,\,\,\,0 & 2.6465800 & 1.4911800 & 0.8623870 \\
 \,\,\,\,\,\,1 & 2.6295900 & 1.4804000 & 0.8551840 \\
2\,\,\,\,\,0 & 1.7526700 & 0.8721010 & 0.4265830 \\
 \,\,\,\,\,\,1 & 1.7394900 & 0.8642480 & 0.4217640 \\
 \,\,\,\,\,\,2 & 1.7133000 & 0.8486670 & 0.4122170 \\
3\,\,\,\,\,0 & 1.0823900 & 0.4428370 & 0.1570130 \\
 \,\,\,\,\,\,1 & 1.0724800 & 0.4374880 & 0.1542230 \\
 \,\,\,\,\,\,2 & 1.0528200 & 0.4269000 & 0.1487240 \\
 \,\,\,\,\,\,3 & 1.0237200 & 0.4112900 & 0.1406770 \\
4\,\,\,\,\,0 & 0.5991580 & 0.1711030 & 0.0242272 \\
 \,\,\,\,\,\,1 & 0.5920830 & 0.1679170 & 0.0231832 \\
 \,\,\,\,\,\,2 & 0.5780760 & 0.1616440 & 0.0211659 \\
 \,\,\,\,\,\,3 & 0.5574210 & 0.1524740 & 0.0183168 \\
 \,\,\,\,\,\,4 & 0.5305400 & 0.1406920 & 0.0148458 \\
5\,\,\,\,\,0 & 0.2734990 & 0.0311495 & 0.0049878 \\
 \,\,\,\,\,\,1 & 0.2689040 & 0.0298494 & 0.0054591 \\
 \,\,\,\,\,\,2 & 0.2598410 & 0.0273349 & 0.0064646 \\
 \,\,\,\,\,\,3 & 0.2465630 & 0.0237765 & 0.0081293 \\
 \,\,\,\,\,\,4 & 0.2294460 & 0.0194276 & 0.0106393 \\
 \,\,\,\,\,\,5 & 0.2089850 & 0.0146225 & 0.0142399 \\
\end{tabular}
\end{ruledtabular}
\end{table}

\newpage

\begin{table}
\begin{ruledtabular}
\caption{Energy eigenvalues of the $LiH$ molecule in $eV$
($E_{n\ell}<0$).}
\begin{tabular}{cccc}
$n\,\,\,\,\,\,\ell$ & $q=1.25$ & $q=1.50$ & $q=1.75$ \\
\hline
0\,\,\,\,\,0 & 3.5677900 & 2.3156300 & 1.5836900 \\
1\,\,\,\,\,0 & 3.0290300 & 1.9055500 & 1.2595300 \\
 \,\,\,\,\,\,1 & 3.0250700 & 1.9029400 & 1.2577000 \\
2\,\,\,\,\,0 & 2.5506900 & 1.5476900 & 0.9820380 \\
 \,\,\,\,\,\,1 & 2.5471400 & 1.5453900 & 0.9804600 \\
 \,\,\,\,\,\,2 & 2.5400700 & 1.5408100 & 0.9773080 \\
3\,\,\,\,\,0 & 2.1274300 & 1.2372900 & 0.7468470 \\
 \,\,\,\,\,\,1 & 2.1242700 & 1.2352900 & 0.7455040 \\
 \,\,\,\,\,\,2 & 2.1179600 & 1.2312900 & 0.7428200 \\
 \,\,\,\,\,\,3 & 2.1085100 & 1.2253000 & 0.7388050 \\
4\,\,\,\,\,0 & 1.7544800 & 0.9701380 & 0.5500770 \\
 \,\,\,\,\,\,1 & 1.7516800 & 0.9684030 & 0.5489500 \\
 \,\,\,\,\,\,2 & 1.7460700 & 0.9649380 & 0.5466990 \\
 \,\,\,\,\,\,3 & 1.7376900 & 0.9597550 & 0.5433330 \\
 \,\,\,\,\,\,4 & 1.7265600 & 0.9528690 & 0.5388640 \\
5\,\,\,\,\,0 & 1.4275700 & 0.7424420 & 0.3882750 \\
 \,\,\,\,\,1 & 1.4251000 & 0.7409570 & 0.3873480 \\
 \,\,\,\,\,2 & 1.4201500 & 0.7379920 & 0.3854990 \\
 \,\,\,\,\,3 & 1.4127600 & 0.7335570 & 0.3827350 \\
 \,\,\,\,\,4 & 1.4029300 & 0.7276680 & 0.3790660 \\
 \,\,\,\,\,5 & 1.3907100 & 0.7203470 & 0.3745090 \\
\end{tabular}
\end{ruledtabular}
\end{table}

\newpage

\begin{table}
\caption{Energy eigenvalues obtained from Eq. (15), Eq. (18) and
Eq. (30).}
\begin{ruledtabular}
\begin{tabular}{ccccccc}
&  &  & Manning-Rosen potential & &  & \\
\hline $n$ & $\ell$ & $1/b$ & & & Present Work & Ref. [19] \\
2 & 1 & 0.025 & & & -0.1205793 & -0.1205279 \\
  &   & 0.050 & & & -0.1084228 & -0.1082170 \\
  &   & 0.075 & & & -0.0969120 & -0.0964490 \\
  &   & 0.100 & & & -0.0860470 & -0.0852240 \\
\hline
&  &  & Hulth\'{e}n potential  & & & \\
\hline $n$ & $\ell$ & $\delta$ & & & Present Work & Ref. [21] \\
0 & 1 & 0.025 & & & -0.1128130 & -0.1127600 \\
  &   & 0.050 & & & -0.1012500 & -0.1010420 \\
  &   & 0.075 & & & -0.0903120 & -0.0898450 \\
  &   & 0.100 & & & -0.0800000 & -0.0791700 \\
  &   & 0.150 & & & -0.0612500 & -0.0594950 \\
\hline
&  &  & Morse potential  & & & \\
\hline $n$ &  &  & & & Present Work & Ref. [17] \\
0  &   &  & & & -4.476013 & -4.476013 \\
1  &   &  & & & -3.962315 & -3.962315 \\
2  &   &  & & & -3.479919 & -3.479918 \\
3  &   &  & & & -3.028824 & -3.028823 \\
4  &   &  & & & -2.609030 & -2.609029 \\
5  &   &  & & & -2.220537 & -2.220536 \\
\end{tabular}
\end{ruledtabular}
\end{table}

\end{document}